# Metadynamics for vacancy dynamics in crystals

Kazuaki Toyoura* and Shunya Yamada

*Department of Materials Science and Engineering, Kyoto University, Kyoto 606-8501, Japan*

*toyoura.kazuaki.5r@kyoto-u.ac.jp

## ABSTRACT

We propose a metadynamics-based (MetaD-based) approach to construct the free energy surface (FES) for vacancy dynamics in crystals. In this approach, the vacancy FES can be constructed without explicitly defining a unique vacancy coordinate by introducing a set of parameters that strongly govern the FES, which is enabled by *parallel bias MetaD with partitioned families* (PB-MetaDPF). In addition, the proposed approach is made more efficient and effective by *multihill strategy* that exploits crystallographic symmetry. We demonstrate the validity and efficacy of the proposed approach by its applications to self-diffusion and impurity diffusion via monovacancies and divacancies in metallic and ionic crystals.

## I. INTRODUCTION

Atomic diffusion in solids plays a key role in many physical phenomena and chemical reactions, and a fundamental understanding at the atomic scale is essential.[1,2] Theoretical calculations are powerful tools for acquiring the microscopic picture of atomic diffusion, among which the molecular dynamics (MD) method[3,4] has been widely used for several decades. However, its applicability is severely limited by the accessible time scales when rare atomic jumps are involved.

A statistical mechanical approach is a solution to overcome such a rare event problem. According to the transition state theory (TST),[5] the atomic jump frequency $\mathit{\Gamma}$ is given by

$$\mathit{\Gamma} = \sqrt{\frac{k_\mathrm{B}T}{2\pi M}}\frac{\int_\mathrm{S} \exp\left(-\frac{U(\mathbf{x})}{k_\mathrm{B}T}\right)d\mathbf{x}^{(3N-1)}}{\int_\mathrm{I} \exp\left(-\frac{U(\mathbf{x})}{k_\mathrm{B}T}\right)d\mathbf{x}^{(3N)}}, \quad (1)$$

where $k_\mathrm{B}$ is the Boltzmann constant, $T$ is the temperature, $M$ is the mass of diffusion species, $N$ is the number of atoms in a system, $\mathbf{x}$ is the coordinate in the $3N$-dimensional configuration space, and $U(\mathbf{x})$ is the potential energy as a function of $\mathbf{x}$. $3N$ and $3N$–1 in the parentheses to the upper right of $\mathbf{x}$ denote the dimension of $\mathbf{x}$, and the integral ranges I and S are the $3N$-dimensional hypervolume corresponding to the initial state for the atomic jump and the $3N$–1-dimensional saddle hypersurface separating the initial and final states, respectively. Under the classical harmonic approximation, the integrations in the denominator and numerator can be performed analytically, resulting in the following simple equation,

$$\mathit{\Gamma}_\mathrm{HA} = \frac{\prod_{i=1}^{3N}\nu_i^\mathrm{I}}{\prod_{i=2}^{3N}\nu_i^\mathrm{S}}\exp\left(-\frac{\Delta U}{k_\mathrm{B}T}\right), \quad (2)$$

where $\Delta U$ is the change in potential energy from the initial point to the saddle point, and $\nu_i^\mathrm{I}$ and $\nu_i^\mathrm{S}$ are the $i$th vibrational eigenfrequencies at the initial and the saddle points, respectively. Note that an imaginary mode at the saddle point ($i$ = 1) is excluded from the denominator in the preexponential factor. In general, the $\Delta U$ and the $\nu_i^\mathrm{I}$ and $\nu_i^\mathrm{S}$ are estimated by the nudged elastic band (NEB) method[6] and the phonon calculations[7], respectively. Once all frequencies of possible atomic jumps in the system of interest are obtained, the atomic diffusivity can readily be estimated by the kinetic Monte Carlo (KMC) method[8] or the numerical solution of master equation[9] without the feasible time scale limitation. However, the NEB-based approach also has a problem that it cannot be applied to dynamically unstable systems, such as finite-temperature phases of perovskites undergoing successive displacive phase transitions.[10] Moreover, the NEB method requires the prior knowledge on the initial and final states, limiting its applicability to unknown diffusion pathways.

The metadynamics (MetaD) method[11-13] is an MD-based enhanced sampling method, which enables the estimation of the atomic jump frequency based on the TST even in dynamically unstable systems. In this method, rare events are accelerated during an MD simulation by sequentially accumulating a small Gaussian bias potential (*Gaussian hill*) on the free energy surface (FES) defined in a few dimensional collective variable (CV) space. The CVs should be chosen to describe the physical phenomenon of interest, e.g., the atomic coordinates of diffusion atoms for atomic transport phenomena. The most important feature of the MetaD method is that it can estimate the FES from the final shape of the bias potential, from which all elementary processes of atomic jumps can be extracted in the system of interest, and their jump frequencies can be estimated based on the TST without the harmonic approximation. In addition, the prior knowledge on the diffusion paths in the system of interest is not required by using *multihill strategy*[14] exploiting the crystallographic symmetry. Thus, the MetaD method has a potential to analyze atomic diffusion in crystals that cannot be evaluated by either MD or NEB method.

A challenge in the MetaD method for atomic diffusion is general treatment for vacancy dynamics in crystals. In most metallic and ionic crystals, atomic diffusion and ionic conduction are known to be mediated by vacancies.[1,15] The specific challenge here is the choice of appropriate CVs describing the dynamics of intangible vacancy. The simplest CV for vacancy dynamics is the relative coordinate of an adjacent diffusion atom to all or a part of the host atoms,[16-18] which is often limited to one dimension along a specified path. However, in the case that another adjacent atom jumps into the vacancy during the simulation, as is often in a system including multiple diffusion paths with different jump frequencies, the original adjacent atom becomes immobile by separating from the vacancy, resulting in the deep free energy well. Although the resulting FES is correct in that it evaluates the tracer-diffusion coefficient of the diffusion species, the MetaD method requires a long time-scale simulation comparable to MD simulations, meaning no use of the acceleration by the MetaD method.

This problem is avoided by defining a unique vacancy coordinate using the coordinates of all adjacent diffusion atoms. The weighted average of the coordinates of all adjacent atoms or adjacent lattice points is sometimes used as the CVs in the literature,[19,20] which seems reasonable as the definition of vacancy coordinate. A concern with these definitions is that the resulting FES drastically changes depending on the setting parameters related to selection of adjacent atoms and their weights. In addition, the vacancy jump frequency cannot be estimated based on Eq. (1) due to the time dependences of the number and weights of adjacent atoms for determining the vacancy coordinate during the simulation.

In the present study, we propose a general and effective MetaD-based approach for vacancy dynamics using the *parallel bias MetaD method with partitioned families* (PB-MetaDPF)[21] combined with the multihill strategy. The key point is that the unique vacancy coordinate is not determined in the proposed approach. Instead, the multiple vacancy coordinates corresponding to individual adjacent atoms are defined, all of which are regarded as the same family in the PB-MetaDPF scheme. Since each vacancy coordinate corresponds to a single diffusion atom, the resulting FES reflects the motion of a virtual particle with the same mass as the diffusion species, meaning that the vacancy jump frequency can be estimated based on Eq. (1). Even in the case of multiple vacancies, we have only to distinguish these vacancies as different partitioned families. The proposed approach is made more efficient and effective by the multihill strategy, which accelerates the Gaussian hill deposition by exploiting the crystallographic symmetry and eliminates the necessity of the prior knowledge on the vacancy diffusion paths in the crystal of interest. As a result, this approach does not require any parameter for definition of the vacancy coordinate, and is applicable to any system with multiple vacancies and multiple diffusion paths. The required information is only the lattice points of diffusion atoms and the coordination number of each lattice point.

In the following, we first describe the theoretical aspects of the proposed approach, and then demonstrate its validity and efficacy using several model cases. Specifically, the monovacancy diffusion in the copper metal with the face-centered cubic crystal structure (fcc-Cu) is first taken as the simplest case, to demonstrate how the proposed approach works. The divacancy diffusion and the impurity indium (In) diffusion in fcc-Cu are also taken as the cases with multiple vacancies and multiple diffusion species, respectively. In addition, the oxygen vacancy diffusion in titanium dioxide with the rutile structure (rutile-$TiO_2$) is taken as the case with multiple diffusion paths in a crystal.

## II. COMPUTATIONAL METHODOLOGY

### A. Proposed approach for vacancy dynamics based on metadynamics

Considering a system containing $n$ diffusion atoms, the CV space describing the diffusion phenomenon has $3n$ dimensions corresponding to the three-dimensional coordinates of all diffusion atoms in a straightforward manner. In the case of the high site occupancy of diffusion species, dilute atomic vacancies can be considered as independent diffusion carriers with negligible interactions. As a result, the high-dimensional CV space ($3n$ dimension) can be reduced down to three-dimensional CV space corresponding to the vacancy coordinate, if it can reasonably be defined as a virtual particle. In the literature,[19,20] the CV vector for vacancy dynamics, $\mathbf{s}_\mathrm{v}$, is related to the coordinates of all atoms adjacent to the vacancy site, $\left\{\mathbf{r}_\mathrm{adj}^{(i)}\right\}$ $(i = 1, 2, \ldots, z) \in \mathbb{R}^{3z}$, where $z$ is the coordination number. The $\mathbf{s}_\mathrm{v}$ in the literature is defined as the three-dimensional vacancy coordinate, $\mathbf{r}_\mathrm{v}$, by some function: $\mathbb{R}^{3z} \longrightarrow \mathbb{R}^3$, i.e., $\mathbf{s}_\mathrm{v} = \mathbf{r}_\mathrm{v}\left(\left\{\mathbf{r}_\mathrm{adj}^{(i)}\right\}\right)$. However, such a function has several parameters concerning the number and weights of adjacent atoms, which drastically influence on the resulting FES.

Instead of determining such a unique vacancy coordinate, the proposed approach explicitly treats the $3z$-dimensional CVs, $\mathbf{s}_\mathrm{v} = \left(\mathbf{s}^{(1)}\left(\mathbf{r}_\mathrm{adj}^{(1)}\right), \mathbf{s}^{(2)}\left(\mathbf{r}_\mathrm{adj}^{(2)}\right), \ldots, \mathbf{s}^{(z)}\left(\mathbf{r}_\mathrm{adj}^{(z)}\right)\right)$, where $\mathbf{s}^{(i)}$ $(i = 1, 2, \ldots, z)$ are the three-dimensional CV vectors associated with the individual adjacent atoms.

Specifically, the vacancy coordinate defined by adjacent atom $i$, $\mathbf{r}_{\mathrm{v}}^{(i)}$, is used as $\mathbf{s}^{(i)}$, i.e., $\mathbf{s}^{(i)} = \mathbf{r}_{\mathrm{v}}^{(i)}(\mathbf{r}_{\mathrm{adj}}^{(i)})$. Figure 1(a) shows the definition of the vacancy coordinate in the proposed approach, $\mathbf{r}_{\mathrm{v}}^{(i)}$, given by

$$\mathbf{r}_{\mathrm{v}}^{(i)} = \mathbf{r}_{\mathrm{v,lat}} - \Delta\mathbf{d}_i, \tag{3}$$

where $\mathbf{r}_{\mathrm{v,lat}}$ is the coordinate of the vacant lattice point. The vacant lattice point is defined as the lattice point that is not a nearest lattice point of any diffusion atom, which can be readily identified simply by measuring the distances between the diffusion atoms and the lattice points. $\Delta\mathbf{d}_i$ is the displacement of atom $i$ from the nearest lattice point, $\Delta\mathbf{d}_i = \mathbf{r}_{\mathrm{adj}}^{(i)} - \mathbf{r}_{\mathrm{adj,latt}}^{(i)}$, where $\mathbf{r}_{\mathrm{adj,latt}}^{(i)}$ is the nearest lattice point to atom $i$. With this definition, the vacancy moves smoothly in the direction opposite to that of the migrating atom during site exchange, although the vacant and the nearest lattice points are exchanged on the Voronoi boundary. This continuity is essential for the accurate evaluation of the FES in the vicinity of the saddle surface. Note that some adjacent atoms except the migrating atom are replaced by other atoms with changing the vacant lattice point. In this case, adjacent atoms included in the original list of adjacent atoms are preferentially assigned to the same CV space as before, while the others are assigned randomly. Refer to Sec. S1 in the Supplementary Material for details regarding the assignment of CV space for each adjacent atom, in which also accounts for cases with multiple vacancies discussed later. Also refer to Sec. S2 in the Supplementary Material for the discussion on guaranteeing the continuities of the total bias potential and the bias forces during the random assignment.

The PB-MetaDPF method can treat a high-dimensional CV vector by dividing it into several low-dimensional CV vectors. In the proposed approach, the 3$z$-dimensional CV vector, $\left(\mathbf{r}_{\mathrm{v}}^{(1)}\left(\mathbf{r}_{\mathrm{adj}}^{(1)}\right), \mathbf{r}_{\mathrm{v}}^{(2)}\left(\mathbf{r}_{\mathrm{adj}}^{(2)}\right), \ldots, \mathbf{r}_{\mathrm{v}}^{(z)}\left(\mathbf{r}_{\mathrm{adj}}^{(z)}\right)\right) \in \mathbb{R}^{3z}$, is divided into $z$ low-dimensional CV spaces defined by individual adjacent atoms, $\mathbf{r}_{\mathrm{v}}^{(i)}\left(\mathbf{r}_{\mathrm{adj}}^{(i)}\right) \in \mathbb{R}^3$ $(i = 1, 2, \ldots, z)$. The $z$ CV vectors are classified into the same partitioned family, since they all are attributed to the same vacant lattice. As a result, all CV vectors are considered identical, and Gaussian hills are deposited in the same three-dimensional CV space to construct the three-dimensional FES of a single vacancy.

In the normal PB-MetaD method,[22] the bias potential in the $i$th CV space after $n$-times Gaussian hill deposition, $V_n^{(i)}\left(\mathbf{s}^{(i)}\right)$, is given by

$$V_n^{(i)}\left(\mathbf{s}^{(i)}\right) = V_{n-1}^{(i)}\left(\mathbf{s}^{(i)}\right) + w_n^{(i)} h_n^{(i)} \exp\left(-\frac{\left\|\mathbf{s}^{(i)}-\mathbf{s}_{\boldsymbol{n}}^{(i)}\right\|^2}{2\left(\sigma^{(i)}\right)^2}\right), \tag{4}$$

where $h_n^{(i)}$ and $\sigma^{(i)}$ are the height and width of the $n$th Gaussian hill, $w_n$ is a conditional weight to account for the correlation effects of the bias potentials deposited in the other CV spaces, and $\mathbf{s}_{\boldsymbol{n}}^{(i)}$ is the current position in the $i$th CV space when the $n$th Gaussian hill is added. $h_n^{(i)}$ is rescaled according to the bias potential at the current position $\mathbf{s}_{\boldsymbol{n}}^{(i)}$ in the well-tempered MetaD manner[12,13] as follows:

$$h_n^{(i)} = h_0 \exp\left(-\frac{V_{n-1}^{(i)}\left(\mathbf{s}_n^{(i)}\right)}{k_\mathrm{B}\Delta T}\right), \tag{5}$$

where $h_0$ is the initial Gaussian height, and $\Delta T$ is a temperature-valued positive parameter determining the decay rate of the Gaussian hill height. $w_n^{(i)}$ is given by

$$w_n^{(i)} = \exp\left(-\frac{V_{n-1}^{(i)}\left(\mathbf{s}_n^{(i)}\right)}{k_\mathrm{B}T}\right) \Big/ \sum_i \exp\left(-\frac{V_{n-1}^{(i)}\left(\mathbf{s}_n^{(i)}\right)}{k_\mathrm{B}T}\right). \tag{6}$$

In the PB-MetaDPF method, degenerate CV spaces are partitioned into families (PFs), in which the bias potentials are merged in the same family as follows:

$$V_n^{\mathrm{PF}(k)}\left(\mathbf{s}^{\mathrm{PF}(k)}\right) = V_{n-1}^{\mathrm{PF}(k)}\left(\mathbf{s}^{\mathrm{PF}(k)}\right) + \sum_{i\in\mathrm{PF}(k)} w_n^{(i)} h_n^{(i)} \exp\left(-\frac{\left\|\mathbf{s}^{\mathrm{PF}(k)}-\mathbf{s}_n^{(i)}\right\|^2}{2\left(\sigma^{(i)}\right)^2}\right), \tag{7}$$

where $V_n^{\mathrm{PF}(k)}$ is the bias potential of the $k$th family after $n$-times Gaussian hill deposition in each CV space, and $\mathbf{s}^{\mathrm{PF}(k)}$ denotes the corresponding CV vector. Note that $V_{n-1}^{(i)}\left(\mathbf{s}_n^{(i)}\right)$ in Eqs. (5) and (6) should be replaced by $V_{n-1}^{\mathrm{PF}(k)}\left(\mathbf{s}_n^{(i)}\right)$ for estimating $h_n^{(i)}$ and $w_n^{(i)}$ in Eq. (7).

For simple monovacancy diffusion, all CVs are assigned to a single family, resulting in a three-dimensional vacancy FES. Owing to crystallographic symmetry, the resulting FES exhibits the same symmetry as the perfect crystal. The multihill strategy therefore improves sampling efficiency by simultaneously depositing Gaussian hills at all crystallographically equivalent points in each CV space. In the PB-MetaDPF with the multihill strategy, a summation over symmetry operations is added to the expression of $V_n^{\mathrm{PF}(k)}$ given by Eq. (7), as follows:

$$V_n^{\mathrm{PF}(k)}\left(\mathbf{s}^{\mathrm{PF}(k)}\right) = V_{n-1}^{\mathrm{PF}(k)}\left(\mathbf{s}^{\mathrm{PF}(k)}\right) + \sum_{i\in\mathrm{PF}(k)}\sum_j w_n^{(i)} h_n^{(i)} \exp\left(-\frac{\left\|\mathbf{s}^{\mathrm{PF}(k)} - \widehat{\mathrm{O}}_{ij}\mathbf{s}_n^{(i)}\right\|^2}{2\left(\sigma^{(i)}\right)^2}\right), \quad (8)$$

where $\widehat{\mathrm{O}}_{ij}$ is the symmetry operation generating *j*th crystallographically equivalent point to $\mathbf{s}_n^{(i)}$. In the case of the supercell consisting of $l\times m\times n$ unit cells with $N_{\mathrm{ep}}$ equivalent points per unit cell, the total number of symmetry operations is $l\times m\times n\times N_{\mathrm{ep}}$, which is a direct indicator of the Gaussian hill deposition efficiency.

Even in the presence of multiple diffusion species, the vacancy FES can be constructed in the same manner by assigning all CVs to a single family. When the site exchange frequency between a monovacancy and each diffusion species is of interest, the CVs are instead partitioned into distinct families according to the diffusion species. In this case, each FES yields the site exchange frequency with the corresponding species. Similarly, for a simulation cell containing multiple vacancies, the CVs should be partitioned according to vacant lattice points. The resulting multiple FESs converge to the same shape in the long-time limit, corresponding to a *single-particle* FES that incorporates the interactions with the other vacancies. Divacancy diffusion without dissociation into monovacancies can also be treated by an appropriate choice of adjacent atoms. Specifically, only adjacent atoms that preserve the divacancy configuration after site exchange are selected for the CV definition. Through appropriate partitioning of the CV spaces into families, the PB-MetaDPF approach is readily applicable to systems containing multiple vacancies and multiple diffusion species. Note that this framework remains applicable even at higher vacancy concentrations, where interactions between vacancies become significant. However, it can reconstruct a single-vacancy FES in the thermal equilibrium state, and therefore cannot be applied to irreversible processes, such as vacancy cluster dissociation.

Once we obtain the three-dimensional vacancy FES, the vacancy jump frequency can be estimated by the following equation,

$$\Gamma = \sqrt{\frac{k_\mathrm{B}T}{2\pi M}} \frac{\iint_\mathrm{S} \exp\left[-\frac{F(\mathbf{R})}{k_\mathrm{B}T}\right] dS}{\iiint_\mathrm{I} \exp\left[-\frac{F(\mathbf{R})}{k_\mathrm{B}T}\right] dV}, \tag{9}$$

where the numerator and denominator correspond to the surface integral over the saddle surface, S, and the volume integral over the vicinity of the initial state, I, respectively. This expression is equivalent to Eq. (1), in which $F(\mathbf{R})$ includes the integration over the remaining $3N$–3 dimensions. The region corresponding to the initial state and the saddle surface are identified using the grid data of the three-dimensional FES as follows:

1. Identify all local minima from the FES grid data simply by comparing the free energy at each grid point with those at all adjacent grid points.
2. Identify the set of grid points belonging to the basin surrounding each local minimum, which corresponding to the integration region of the initial state.
3. For any pair of local minima, check if the intersection of their corresponding grid point sets is non-empty. A non-empty intersection indicates the existence of a vacancy jump pathway, and the intersection itself constitutes the saddle surface.

The surface and volume integrals were performed using the Monte Carlo integration. The Monte Carlo sampling continues until the integral values converge sufficiently.

## B. Computational Conditions

All calculations in the present study were based on the density functional theory (DFT) using the projector augmented wave (PAW) method[23,24] as implemented in the Vienna ab initio simulation package (VASP) code[25,26]. The generalized gradient approximation parametrized by Perdew, Burke, and Ernzerhof (PBE GGA) was employed for the exchange-correlation functional.[27] In the PAW datasets included the following valence configurations: $3d^{10}4p^1$ for Cu, $4d^{10}5s^25p^1$ for In, $3d^34s^1$ for Ti, and $2s^22p^4$ for O. Note that $3d^{10}4p^1$ for Cu represents the reference atomic state used in the generation of the VASP Cu PAW pseudopotential. The plane-wave cutoff energies were set to 320 eV for fcc-Cu and 500 eV for rutile-$TiO_2$. Brillouin-zone integrations for the unit cells were performed using Monkhorst–Pack meshes of 12×12×12 for fcc-Cu and 4×4×6 for rutile-$TiO_2$.

For phonon calculations and MetaD simulations, supercells consisting of 2×2×2 fcc-Cu unit cells and 2×2×3 rutile-$TiO_2$ unit cells were employed, with corresponding *k*-point meshes of 6×6×6 and 2×2×2. Defective supercells containing a monovacancy, a divacancy, and/or an impurity atom were constructed by appropriately adding or removing atoms and electrons. When required, a uniform background charge was introduced to ensure the charge neutrality. Prior to the phonon calculations and MetaD simulations, all defective structures were fully relaxed until the residual forces on all atoms were smaller than $1\times10^{-5}$ eV/Å. Phonon calculations for fcc-Cu were performed using the finite-displacement method as implemented in the PHONOPY code[28] to evaluate the vibrational eigenfrequencies in both perfect and defective supercells.

Before the MetaD simulations, NPT-ensemble molecular dynamics (MD) simulations were performed for 30 ps for perfect supercells of fcc-Cu and rutile-$TiO_2$ to determine the equilibrium lattice constants at finite temperatures. The temperature and pressure were controlled using the Langevin thermostat[29] and the Parinello-Rahman barostat[30], respectively. The time step was set to 1 fs for all MD and subsequent MetaD simulations. The first 10 ps were discarded as equilibration, and lattice constants were obtained as time averages over the remaining 20 ps. Using these lattice constants, additional NVT-MD simulations were performed for 10 ps for the defective supercells as further equilibration steps. Subsequently, NVT-ensemble well-tempered PB-MetaDPF simulations with the multihill strategy were performed according to our proposed approach. For the Cu monovacancy and divacancy diffusions in fcc-Cu with or without an In impurity, the crystallographically equivalent points were generated by the symmetry operations attributed to the space group of $Fm\bar{3}m$ (No. 225) and the translational operations attributed to the 2×2×2 supercell. For the oxygen vacancy diffusion in rutile-$TiO_2$, the equivalent points were generated by the symmetry operations attributed to the space group of $P4_2/mnm$ (No. 136) and the translational operations attributed to the 2×2×3 supercell. Note that the symmetry is overestimated for multiple-vacancy diffusion or vacancy diffusion with impurities in a rigorous manner, although these symmetry operations based on the perfect crystal are appropriate for the monovacancy diffusions in fcc-Cu and rutile-$TiO_2$, Therefore, the FES obtained in the current manner is an averaged FES over all possible local environments. To obtain a preliminary estimate of the free-energy surface (FES), rough PB-MetaDPF simulations with large Gaussian hills were first conducted for 20 ps. The initial height $h_0$ and width $\sigma$ of Gaussian hills and the parameter $k_B\Delta T$ were set to 0.02 eV, 0.2 Å, and 0.4 eV for fcc-Cu, and 0.03 eV, 0.3 Å, and 0.5 eV for rutile-$TiO_2$, respectively. Gaussian hills were deposited every 100 fs. Accurate PB-MetaDPF simulations were then performed for 100 ps using smaller Gaussian hills, where $h_0$, $\sigma$, and $k_B\Delta T$ were 0.01 eV, 0.15 Å, and 0.2 eV for fcc-Cu and 0.02 eV, 0.2 Å, and 0.4 eV for rutile-$TiO_2$, respectively, with hill deposition every 50 fs. The accurate PB-MetaDPF simulations were repeated eight times at each temperature to statistically estimate the vacancy jump frequencies and diffusivities. The simulation temperatures ranged from 500 K to 1300 K for fcc-Cu and from 500 K to 1000 K for rutile-$TiO_2$, in increments of 100 K. See Fig. S2 in Sec. S3 for the convergence tests for the several parameters, i.e., the initial height $h_0$, width $\sigma$, and deposition interval of Gaussian hills.

For the MD and MetaD simulations, on-the-fly machine learning force fields (MLFFs)[31-33] implemented in the VASP code were employed. The MLFFs were based on the smooth overlap of atomic positions (SOAP) descriptor,[34] and were trained during the MD and rough PB-MetaDPF simulations using the on-the-fly learning scheme. The constructed MLFFs were subsequently employed for the accurate PB-MetaDPF simulations. Their accuracy was assessed by comparing total energies and atomic forces obtained from MLFFs and DFT calculations for several hundred structures sampled during the accurate PB-MetaDPF simulations. As shown in Figs. S3 and S4 of Sec. S4 in the supplementary material, the MLFFs reproduce the DFT results with sufficient accuracy.

## III. RESULTS AND DISCUSSION

### A. Monovacancy diffusion in fcc-Cu

Monovacancy diffusion in fcc-Cu is taken as the first example to demonstrate the proposed approach. In fcc-Cu, a monovacancy is surrounded by twelve first-nearest-neighbor (1NN) Cu sites, and twelve vacancy coordinates corresponding to the adjacent atoms are defined, as shown in Fig. 1(b). Figure 1(c) shows the vacancy FES constructed from a single PB-MetaDPF run at 500 K. Owing to the multihill strategy, the constructed FES accurately reflects the crystallographic symmetry of the fcc lattice. According to the FES, the vacancy basically resides on the Cu site and migrates to a 1NN Cu site along one of twelve equivalent diffusion paths. Figure 2 shows the profiles of vacancy jump counts and the free energy barriers in the PB-MetaD simulations at 500, 700, 900, 1100, and 1300 K for 100 ps. The vacancy jump count increases with increasing temperature, while the convergence of the free energy barrier is comparable regardless of temperature. This is attributed to the fact that the multihill strategy fills all free energy basins simultaneously in accordance with the crystallographic symmetry. Consequently, the state transitions of the vacancy are not as critical for the FES evaluation as those in the conventional MetaD, leading to no dependence of the free energy barrier profile on the vacancy jump frequency. The red symbols in Fig. 3(a) show the vacancy jump frequencies $\Gamma_{\mathrm{V}}$ estimated from Eq. (9) as a function of inverse temperature in the range of 500 – 1300 K. At each temperature, eight independent PB-MetaDPF runs were performed, and the error bars represent $\pm\sigma$, where $\sigma$ is the standard deviation among the runs. The preexponential factor $\Gamma_0$ and the apparent activation energy $Q$ are $1.7\times10^{12}$ /s and 0.81 eV, respectively. The gray symbols denote the jump frequencies estimated by the conventional MetaD simulations using the coordinate of a single adjacent Cu atom as the CVs, during which another adjacent Cu atom is reselected if the vacancy moves away from the original adjacent Cu atom. Note that this conventional MetaD method with the simple CV definition is made possible by the multihill strategy only when all adjacent atoms are crystallographically equivalent. The red line is in excellent agreement with the gray one ($\Gamma_0 = 1.6\times10^{12}$ /s and $Q = 0.82$ eV), indicating that a FES constructed from multiple vacancy coordinates in the proposed approach is equivalent to a single vacancy FES. For comparison, the black broken line shows the jump frequencies estimated according to Eq. (2) under the harmonic approximation ($\Gamma_0 = 4.4\times10^{12}$ /s and $Q = 0.75$ eV), where the vibrational eigenfrequencies at the initial and saddle points are shown in Fig. S5(a) of Sec. S5 in the supplementary material. The harmonic approximation overestimates the preexponential factor and underestimates the apparent activation energy, leading to the overestimation of vacancy jump frequency in fcc-Cu. See Sec. S6 for the temperature dependence in the one-dimensional FES along the migration coordinate, in which the difference from the harmonic approximation is discussed.

**B. Divacancy diffusion in fcc-Cu**

The divacancy diffusion in fcc-Cu is the second example, in which we consider only the Cu jumps not separating the two adjacent vacancies. Specifically, the coordinates of four Cu atoms adjacent to both vacancies were employed for defining the vacancy coordinates, meaning that there are four vacancy coordinates per vacant lattice. The four coordinates attributed to a vacant lattice are assigned in the same family, while the CVs attributed to the other vacant lattice are partitioned as a distinct family. Figure 1(d) shows the vacancy FES via divacancy constructed from a single run at 500 K, which is totally different from that via monovacancy shown in Fig. 1(c). In the FES associated with divacancy diffusion, a metastable state emerges near the center of the triangle formed by three adjacent Cu sites. An adjacent Cu atom does not directly jump into the divacancy, but migrates via the metastable site. The corresponding free energy barrier is lower than that in the monovacancy FES by ~0.2 eV. The blue symbols in Fig. 3(a) show the estimated vacancy jump frequencies via divacancy as a function of inverse temperature in the range of 500 – 1100 K. The estimated FESs and jump frequencies at 1200 and 1300 K are inaccurate due to the divacancy dissociation, which are therefore not shown in the figure. $\Gamma_0$ and $Q$ of the jump frequencies via divacancy are $7.7\times10^{11}$ /s and 0.56 eV, respectively, leading to the higher jump frequency relative to that via monovacancy.

Based on the vacancy jump frequencies, the Cu tracer diffusion coefficients $D^*_{\mathrm{Cu}}$ via monovacancies and divacancies were estimated using the conventional equation, $D^* = f x_{\mathrm{v}} D_{\mathrm{v}}$, where $f$, $x_v$, and $D_v$ are the correlation factor, the vacancy site fraction, and the vacancy diffusion coefficient, respectively. The correlation factors of monovacancy and divacancy diffusions, $f_v$ and $f_{2v}$, are reported to be 0.7815 and 0.4582 in the fcc lattice, respectively.[1,35] The site fractions of monovacancies and divacancies, $x_v$ and $x_{2v}$, were estimated from the formation free energy under the harmonic approximation, which are shown in Fig. S5(b). For $D^*_{Cu}$ via divacancies, an additional factor, 2/3, is also necessary considering the difference in the number of Cu jump paths between monovacancy and divacancy diffusions.[36] Figure 3(b) shows the estimated $D^*_{Cu}$ via monovacancies and divacancies, in which the experimental values in the literature[37] are also shown for comparison. The estimated $D^*_{Cu}$ via monovacancies is in good agreement with the experimentally reported values (red broken line), while that via divacancies is slightly underestimated with respect to the experimentally reported values (blue broken line). The preexponential factor and activation energy of the estimated $D^*_{Cu}$ via monovacancies are $2.9\times10^{-2}$ cm$^2$/s and 1.93 eV, respectively, both of which are slightly underestimated in comparison with the experimentally-reported values, $16\times10^{-2}$ cm$^2$/s and 2.07 eV.[37] Concerning $D^*_{Cu}$ via divacancies, the estimated preexponential factor and the activation energy are 0.85 cm$^2$/s and 2.70 eV, vs. 6.4 cm$^2$/s and 2.59 eV in the literature.[37] Note that the divacancy diffusivity in the literature was roughly estimated from the slight bending of the Arrhenius plot around the melting point (1358 K) by assuming that the bending is attributed only to the divacancy diffusion.

**C. In impurity diffusion in fcc-Cu**

The proposed approach can also be applied to impurity diffusion, taking the dilute In diffusion in fcc-Cu as a model case. In the literature,[38] the site exchange frequency of a vacancy with an In atom, $\Gamma_{v,In}$, is approximately one order of magnitude higher than that with a Cu atom, $\Gamma_{v,Cu}$, around 1000 K in fcc-Cu. In addition, dilute In impurities are known to enhance the Cu diffusivity by a few times in the same temperature range. Here, we examine the reproducibility of these effects by the proposed approach.

In the PB-MetaDPF simulations, a supercell containing a Cu vacancy and an adjacent In atom was employed as the initial configuration. Twelve vacancy coordinates corresponding to the individual adjacent atoms were used as the CVs, and the CVs associated with the adjacent In atom were partitioned into a distinct family. This procedure yielded two types of FESs corresponding to site exchange with a Cu atom and with an In atom. Figures 4(a) and (b) show the resulting FESs at 1000 K obtained from a single PB-MetaDPF simulation. The free energy barrier of site exchange with a Cu atom is comparable to that in pure Cu shown in Fig. 1(c), whereas that of site exchange with an In atom is less than half of this value. $\Gamma_{\mathrm{v,Cu}}$ and $\Gamma_{\mathrm{v,In}}$ estimated from the constructed FESs as a function of inverse temperature are shown in Fig. 4(c). Reflecting the difference in the free energy barrier height, the two apparent activation energies are 0.79 eV for $\Gamma_{\mathrm{v,Cu}}$ and 0.33 eV for $\Gamma_{\mathrm{v,In}}$, leading to significantly higher $\Gamma_{\mathrm{v,In}}$. Above 1000 K, $\Gamma_{\mathrm{v,In}}$ exceeds $\Gamma_{\mathrm{v,Cu}}$ by several tens of times. Furthermore, $\Gamma_{\mathrm{v,Cu}}$ in the presence of an In impurity is slightly higher than that in pure Cu. These trends are consistent with previously reported experimental observations.[38]

**D. O vacancy diffusion in rutile-$TiO_2$**

The oxygen vacancy diffusion in rutile-$TiO_2$ is the final example for application of the proposed approach, in which three types of possible diffusion paths, paths A, B, and C, are considered as shown in Fig. 5(a). According to the literature,[39] the potential energy barriers of paths A, B, and C are reported to be 1.77 eV, 0.69 eV, and 1.10 eV, respectively, evaluated by the NEB method based on DFT calculations. The free energy barrier for path C is also reported, 1.22 eV[40] and 1.44 eV[19] evaluated in different manners. Figure 5(b) shows the two cross-sectional 2D-FESs on planes I and II at 500 K, which contain paths A & C and path B, respectively. According to the constructed FES, the minimum free energy paths are identified along paths B and C, whereas no such path is observed along path A. The free energy barriers along paths B and C are ~0.9 eV and ~1.2 eV, respectively, which slightly varied across eight PB-MetaDPF runs. The FES qualitatively agree with the reported potential energy barriers of paths A, B, and C in terms of the order of barrier heights, but differs from the reported PES in that there is no direct route along path A. Concerning path C, the obtained free energy barrier coincides with that reported in Ref. 40.

Figure 5(c) shows the estimated jump frequencies of oxygen vacancy along paths B and C as a function of inverse temperature. The preexponential factors and the apparent activation energies are $1.3\times10^{13}$/s and 0.96 eV for path B and $9.4\times10^{11}$/s and 1.13 eV for path C, respectively. The apparent activation energies roughly correspond to the free energy barriers, and the jump frequency along path B is two or three orders of magnitude higher than that along path C reflecting the difference in the free energy barrier in this temperature range of 500 – 1000 K. By contrast, the vacancy diffusion coefficient estimated from these jump frequencies exhibits small anisotropy as shown in Fig. 5(d). The apparent activation energies in the ab-plane and along the c-axis are the same, 1.13 eV, while the preexponential factors are slightly different, $2.0\times10^{-3}$ cm$^2$/s in the ab-plane vs. $8.7\times10^{-4}$ cm$^2$/s along the c-axis. The apparent activation energies of the vacancy diffusion coefficients are equal to that of the jump frequency along the c-axis, indicating that long-range migration of oxygen vacancies cannot proceed solely via path B and that path C constitutes the rate-determining process for the oxygen vacancy diffusion in rutile-$TiO_2$. Note that the proposed approach does not require pre-specifying a diffusion path, unlike the NEB method and the conventional MetaD method, which is due to a major advantage of the multihill strategy exploiting the crystallographic symmetry.

## IV. CONCLUSIONS

We have proposed a general and effective MetaD-based approach for vacancy dynamics by combining the PB-MetaDPF method with the multihill strategy. By appropriate CV selection and partitioning, the proposed approach is applicable not only to monovacancy diffusion but also to divacancy diffusion and impurity diffusion mediated by vacancies. Its validity and efficacy were demonstrated for self-diffusion and In impurity diffusion via monovacancies and divacancies in fcc-Cu, as well as for oxygen vacancy diffusion in rutile-$TiO_2$. The present approach provides a general framework for analyzing vacancy-mediated atomic jumps in crystals where conventional MD and NEB-based methods are difficult to apply, and is therefore expected to be a useful tool for studying diffusion phenomena in a wide range of crystalline materials. The obtained FESs by the MetaD-based approach enable us to estimate the vacancy jump frequency based on the TST, which naturally incorporates the anharmonicity of lattice dynamics. For a more rigorous kinetic evaluation beyond the standard TST, the FES obtained herein can serve as a robust basis for employing additional advanced techniques to accurately estimate transition rates, properly accounting for dynamical effects such as concerted jumps and barrier recrossing.

## SUPPLEMENTARY MATERIAL

See the supplemental material for the pseudocode of the proposed approach for vacancy dynamics (Sec. S1), discussion of the continuities of the total bias potential and bias forces (Sec. S2), the convergence tests of MetaD simulations for setting parameters (Sec. S3) the accuracies of the constructed MLFFs (Sec. S4), the phonon calculations in fcc-Cu (Sec. S5), and one-dimensional FESs for Cu monovacancy diffusion at 500 K (Sec. S6). The FES data and the scripts for jump path search and jump frequency estimation are also available.

## ACKNOWLEDGMENTS

I gratefully acknowledge the discussion with K. Tsuru. This work was supported by Precursory Research for Embryonic Science and Technology (PRESTO, Grant No. JPMJPR24J8) from Japan Science and Technology Agency (JST), and KAKENHI (Grants No. 24K01147) from Japan Society for the Promotion of Science (JSPS).

## AUTHOR DECLARATIONS

### Conflict of Interest

The authors have no conflicts to disclose.

### Author Contributions

**Kazuaki Toyoura:** Conceptualization (lead); Formal analysis (lead); Funding acquisition (lead); Investigation (lead); Methodology (lead); Resources (lead); Software (lead); Validation (lead); Visualization (lead); Writing – original draft (lead); Writing – review & editing (lead)

**Shunya Yamada:** Investigation (supporting); Methodology (supporting); Software (supporting); Writing – original draft (supporting); Writing – review & editing (supporting)

## DATA AVAILABILITY

The FES data in the CHGCAR format of VASP code and the scripts for jump path search and jump frequency estimation are available from the supplementary material. The other data that support the findings of this study are available from the corresponding author upon reasonable request.

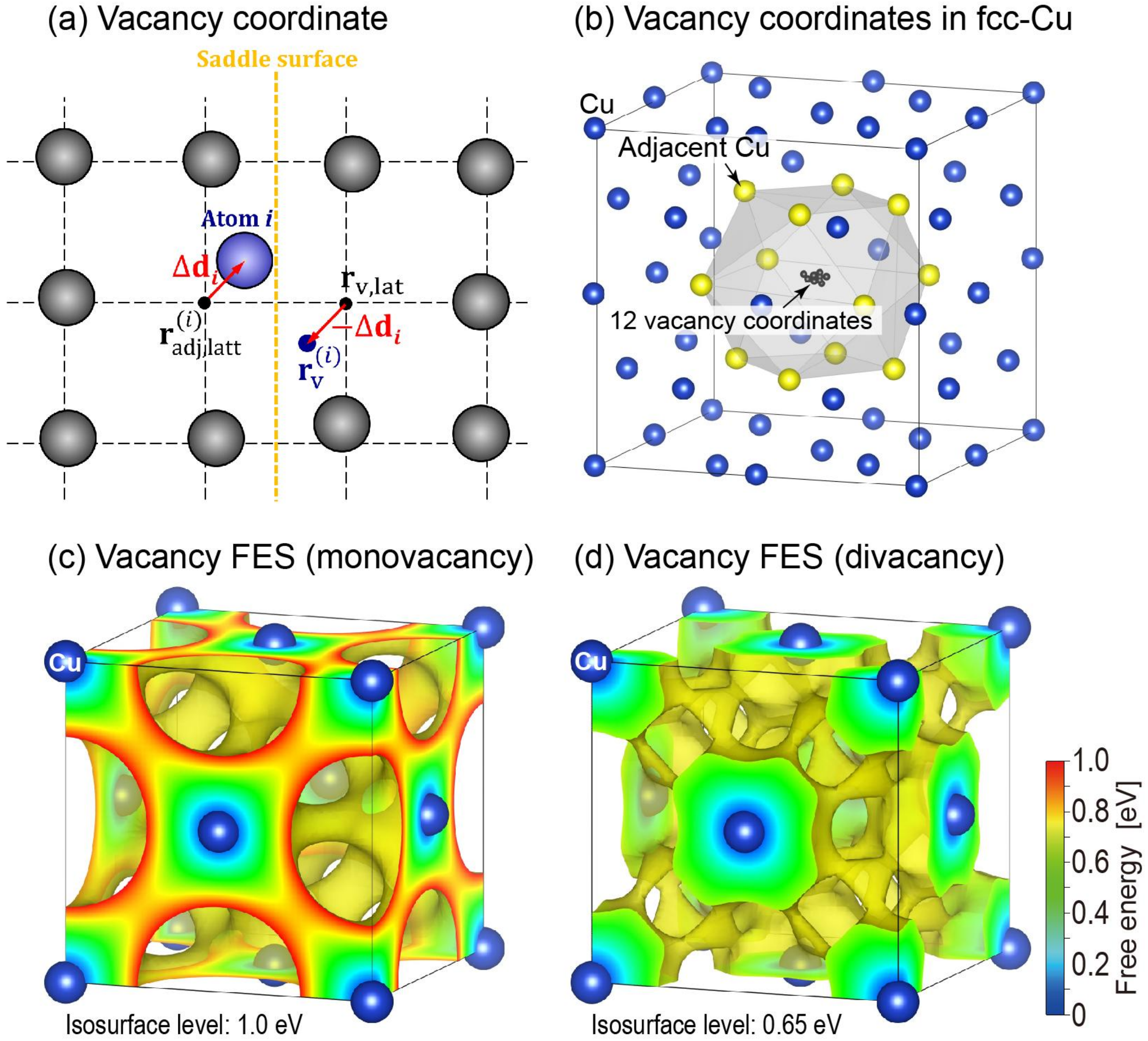

FIG. 1. (a) The vacancy coordinate (blue dot) defined from an adjacent diffusion atom (blue sphere). (b) Twelve vacancy coordinates (black spheres) defined from the individual twelve adjacent atoms (yellow spheres) in fcc-Cu. The vacancy FESs for (c) monovacancy and (d) divacancy diffusion in fcc-Cu constructed from single PB-MetaDPF runs at 500 K. Yellow surfaces denote isosurfaces of the constructed FESs (isosurface levels: (c) 1.0 eV and (d) 0.65 eV).

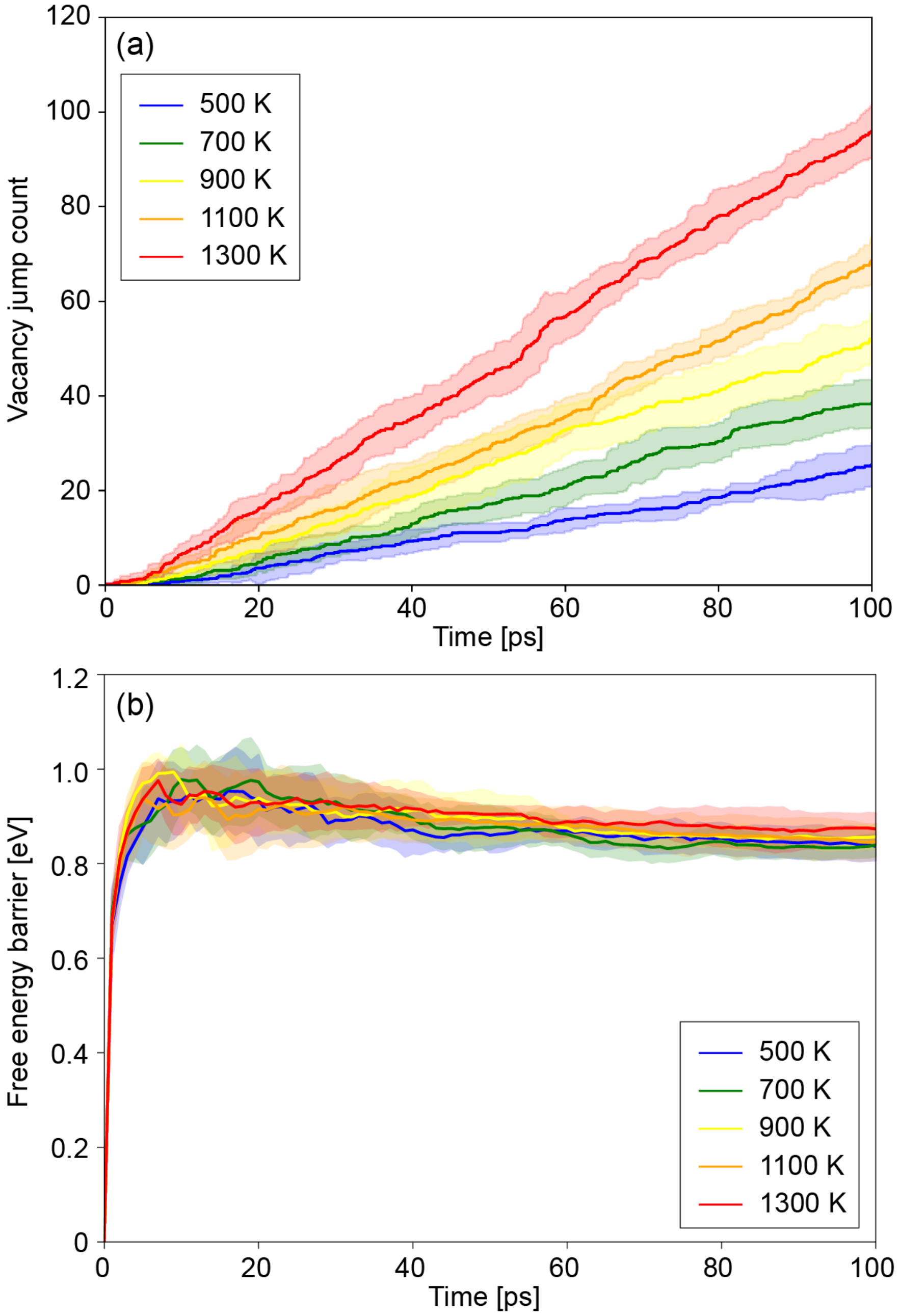

FIG. 2. The profiles of (a) the vacancy jump counts and (b) the free energy barriers in the PB-MetaD simulations for the Cu monovacancy diffusion at 500, 700, 900, 1100, and 1300 K for 100 ps. The solid lines and shaded regions denote the mean and $\pm\sigma$ over eight independent runs at each temperature, respectively.

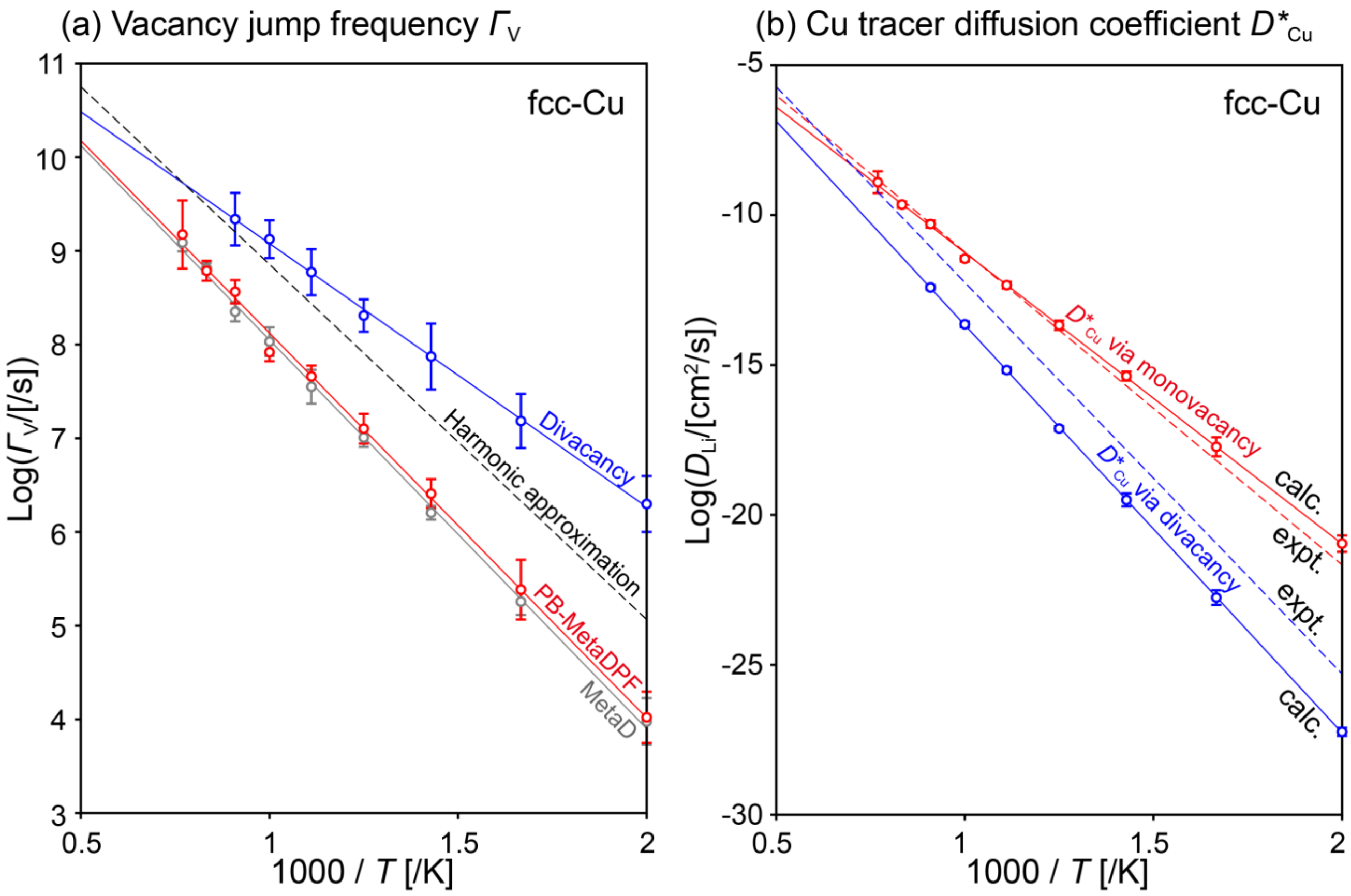

FIG. 3. (a) Arrhenius plots of the vacancy jump frequencies via monovacancy estimated from the constructed FESs by the PB-MetaDPF method (red symbols) and the conventional MetaD method (gray symbols), and those via divacancy (blue symbols). The black broken line denotes the vacancy jump frequencies estimated under the harmonic approximation according to Eq. (2). (b) Arrhenius plots of the estimated Cu tracer diffusion coefficients $D^*_{Cu}$ via monovacancy (red symbols) and divacancy (blue symbols) from the estimated jump frequencies by the PB-MetaDPF method. The red and blue broken lines denote the experimental $D^*_{Cu}$ via monovacancy and divacancy, respectively, in the literature [37]. The error bars in both figures denote $\pm\sigma$ over eight MetaD runs at each temperature.

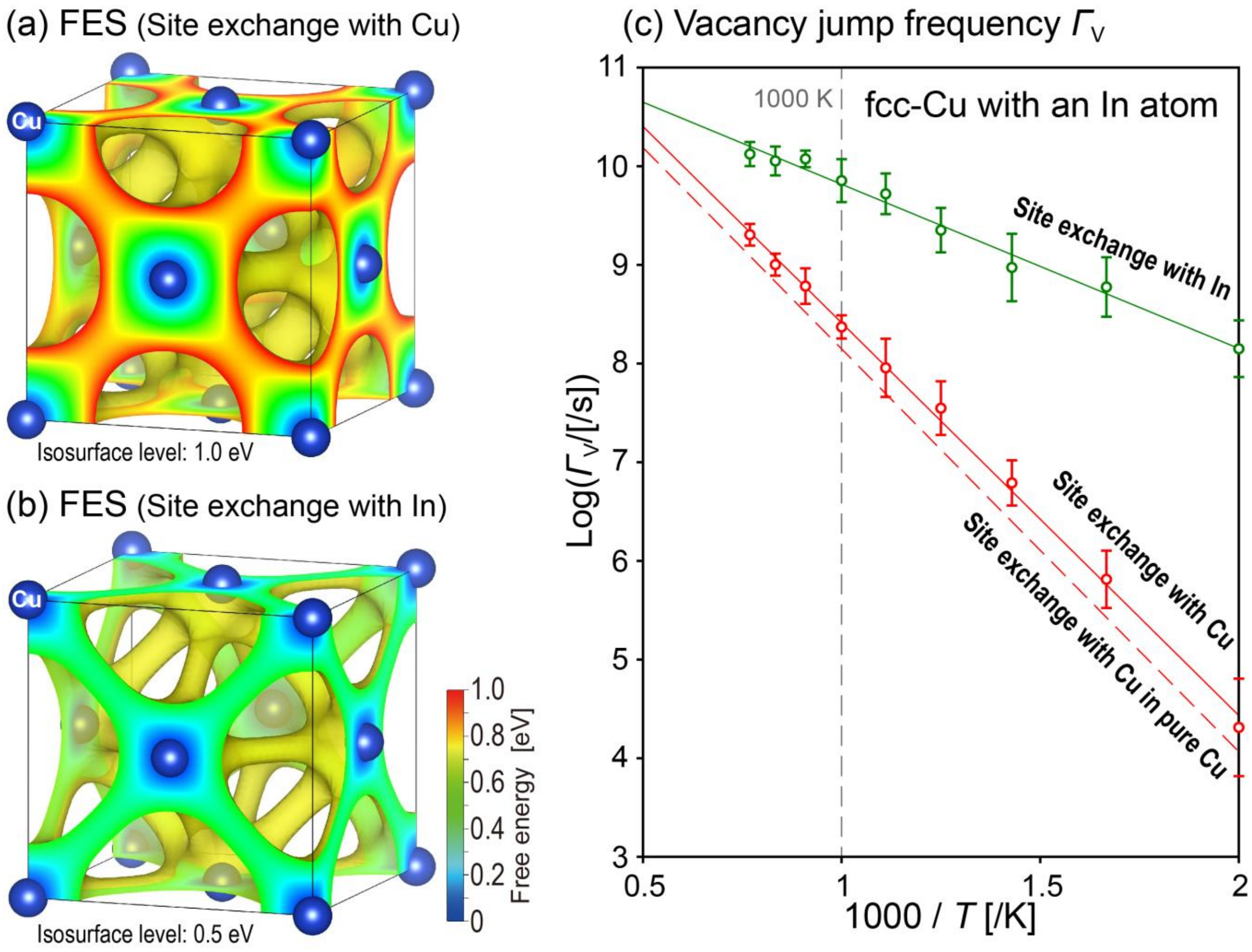

FIG. 4. Vacancy FESs for site exchange with (a) a Cu atom and (b) an In impurity atom in fcc-Cu, constructed from a single PB-MetaDPF run at 1000 K. Yellow surfaces denote isosurfaces of the constructed FESs (isosurface levels: (a) 1.0 eV, (b) 0.5 eV). (c) Arrhenius plots of vacancy jump frequencies for site exchange with a Cu atom (red symbols) and an In atom (green symbols), estimated from the constructed FESs using the PB-MetaDPF method. The error bars denote $\pm\sigma$ over eight MetaD runs at each temperature. The red broken line denotes the vacancy jump frequency in pure Cu shown by the red symbols in Fig. 3(a).

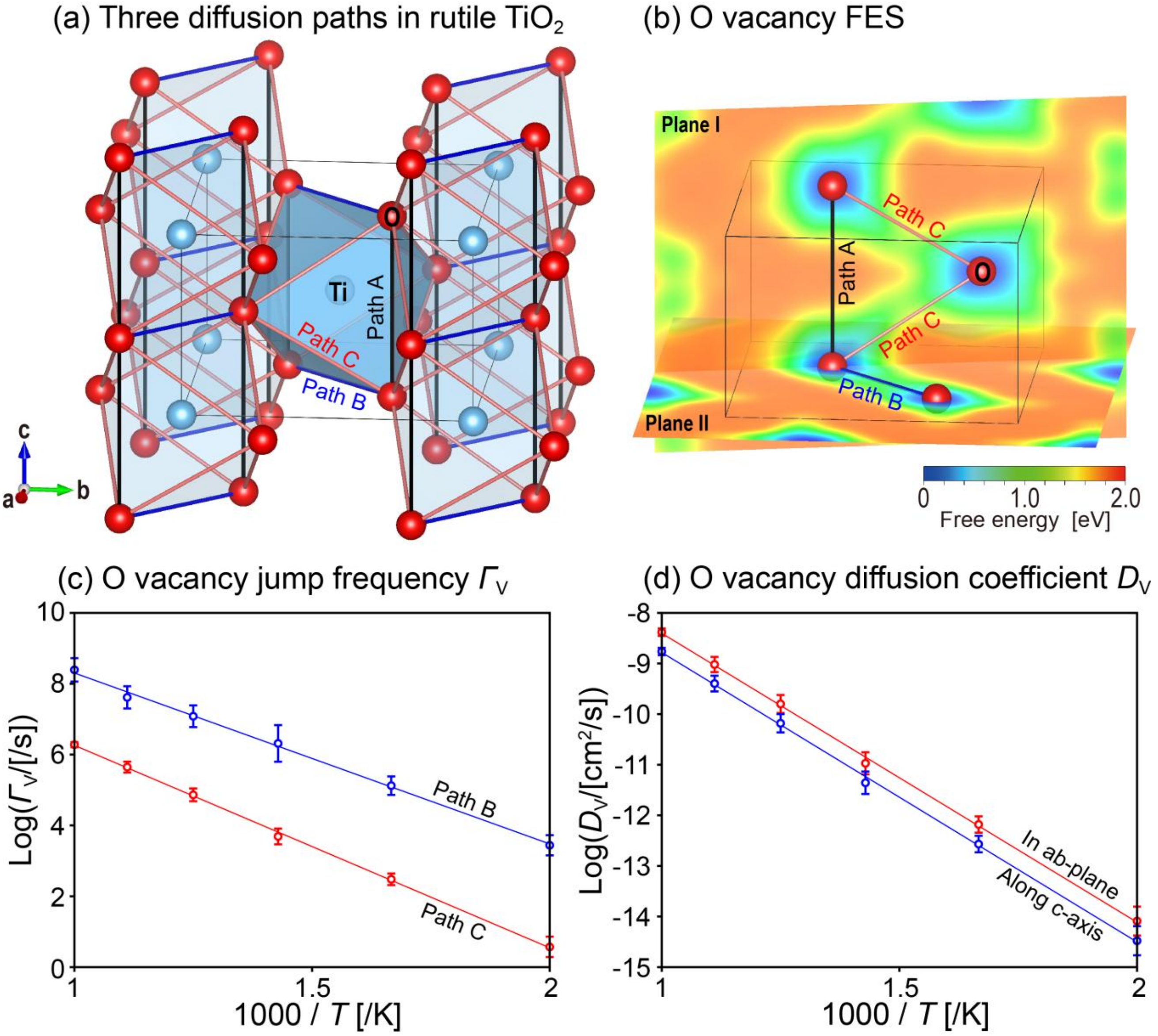

FIG. 5. (a) Three possible diffusion paths of an O vacancy along the edges of $TiO_6$ octahedra (path A: black line, path B: blue line, path C: red line). (b) The two cross-sectional 2D-FESs on planes I and II, which contain paths A & C and path B, respectively, constructed from single PB-MetaDPF runs at 500 K. (c) Arrhenius plots of estimated jump frequencies of oxygen vacancy along paths B and C. (d) Arrhenius plots of estimated diffusion coefficients of oxygen vacancy in the ab-plane and along the c-axis. The error bars in figures (c) and (d) denote $\pm\sigma$ over eight MetaD runs at each temperature.

## *SUPPLEMENTARY MATERIAL*

# Metadynamics for vacancy dynamics in crystals

Kazuaki Toyoura* and Shunya Yamada

*Department of Materials Science and Engineering, Kyoto University, Kyoto 606-8501, Japan*

## S1. Pseudocode of the proposed approach for vacancy dynamics

Table S1 shows the pseudocode of the proposed approach based on the PB-MetaDPF with the multihill strategy, which provides, in particular, the details of the PF and CV mappings.

TABLE S1. Pseudocode of the proposed approach for vacancy dynamics.

```
Algorithm: Dynamic Vacancy CV Evaluation and PB-MetaDPF Simulation
Input:
R_lattice: Reference lattice points
R_atoms: Atomic coordinates at a step
Initialization:
  V_prev ← R_lattice \ { argmin R∈R_lattice ||r – R|| for each r ∈ R_atoms } // Identify vacancies
  Assign unique PF_id and CV_id to each vacant lattice point v ∈ V_prev
Main Loop (PB-MetaDPF):
for step = 1 to N_steps do
  V_curr ← R_lattice \ { argmin R∈R_lattice ||r – R|| for each r ∈ R_atoms } // Identify vacancies
  // Update PF and CV mappings only if vacancy topology changes
  if V_curr ≠ V_prev then
    V_persistent ← V_curr ∩ V_prev
    V_new ← V_curr \ V_prev
    PF_unassigned ← All_PFs \ { PF_id of v | v ∈ V_persistent }
    // Assign PFs: Retain persistent, assign new via adjacency or random fallback
    for each v_new ∈ V_new do
      pf_candidates ← FindAdjacentVacantPFs(v_new) ∩ PF_unassigned
      pf_selected ← RandomChoice(pf_candidates) if non-empty else RandomChoice(PF_unassigned)
      Assign pf_selected to v_new
      PF_unassigned ← PF_unassigned \ {pf_selected}
    end for
    // Assign CVs within each PF: Prioritize persistent CVs, assign remaining randomly
    for each pf ∈ All_PFs do
      UpdateCVAssignmentsWithinPF(pf, V_persistent, V_new)
    end for
  end if
  // Calculate CVs and execute PB-MetaDPF simulation step
  CV_values ← { CalculateVacancyCV(FindNeighbors(v, R_atoms)) for each v ∈ V_curr }
  Execute_PBMetaDPF_Step(CV_values)
  V_prev ← V_curr
end for
```

Alt text: In the proposed approach, the PFs and CVs are changed discontinuously when a vacancy migrate to another lattice site. Because this pseudocode contains the complete mathematical and operational logic, any reader can readily implement this method in their preferred language or integrate it into open-source MD frameworks.

**S2. Continuities of the total bias potential and bias forces**

In our proposed approach, some CV values assigned to newly selected adjacent atoms change discontinuously when a migrating atom crosses a Voronoi boundary. In this section, we provide a mathematical argument demonstrating that PB-MetaD with the multi-hill strategy inherently guarantees the continuities of both total bias potential and bias forces, and demonstrate these profiles before and after the CV reassignment as numerical evidence taking the Cu monovacancy diffusion as an example.

In the PB-MetaD framework, the total bias potential $V_{\text{total}}^{\text{PF}}$ is defined as

$$V_{\text{total}}^{\text{PF}} = -k_{\text{B}}T\ln\left[\sum_i \exp\left(-\frac{V^{(i)}}{k_{\text{B}}T}\right)\right],$$

where $k_{\text{B}}$ is the Boltzmann constant, $T$ is the temperature, and $V^{(i)}$ is the bias potential in the $i$th CV space. Consequently, the bias force acting on atom $i$, $F^{(i)}$, is given by

$$F^{(i)} = -\nabla_i V_{\text{total}}^{\text{PF}} = \left[\exp\left(-\frac{V^{(i)}}{k_{\text{B}}T}\right)\Big/\sum_i \exp\left(-\frac{V^{(i)}}{k_{\text{B}}T}\right)\right] \times \left[-\nabla_i V^{(i)}\right] = -w^{(i)}\nabla_i V^{(i)},$$

where $w^{(i)}$ is the exponential weight for atom $i$. This construction provides a natural and rigorous filtering mechanism against discontinuities during CV reassignment due to the following reasons.

The update of the adjacent-atom list occurs only when an adjacent atom crosses a Voronoi boundary. At the list update, the adjacent atom (migrating atom) is in the transition state, while the other adjacent atoms remain stable in the vicinity of the lattice points. In the case of standard PB-MetaD, the bias potentials $V^{(i)}$ in the CV spaces reassigned to the atoms newly added to the adjacent-atom list are likely to switch discontinuously to zero in the initial stages of the simulation. By contrast, a key feature of the proposed approach is the use of multihill strategy, which fills up all the free energy basins at the same time. Consequently, it is guaranteed that the $V^{(i)}$ values for all other adjacent atoms except the migrating atom remain relatively large before and after the update of the adjacent-atom list, leading to the dominant exponential term of the migrating atom in the summation in the equations of both $V_{\text{total}}^{\text{PF}}$ and $F^{(i)}$.

## *SUPPLEMENTARY MATERIAL*

The $V_{\mathrm{total}}^{\mathrm{PF}}$ profile is almost equal to the $V^{(i)}$ profile of the migrating atom during a certain period before and after the list update. In our proposed approach, the CV assignment for the migrating atom remains unchanged, leading to the smooth $V^{(i)}$ and $V_{\mathrm{total}}^{\mathrm{PF}}$ profiles. Panels (a) and (b) in Fig. S1 show the $V_{\mathrm{total}}^{\mathrm{PF}}$ and $V^{(i)}$ profiles during ±500 fs and ±100 fs with respect to the adjacent-atom list update ($t$ = 0) in a PB-MetaDPF simulation for a single Cu monovacancy at 500 K. In the regions far from the time when the adjacent-atom list is updated ($t$ ~ ±500 ps), the total bias potential reflects the bias potentials $V^{(i)}$ of all adjacent atoms, because all adjacent atoms are stable in the vicinity of the lattice points, i.e., comparable $V^{(i)}$ values for all CV spaces. By contrast, as an atom approaches the transition state near a Voronoi boundary, the total bias potential begins to coincide with the single $V^{(i)}$ profile of the migrating atom. Focusing on panel (b), the $V_{\mathrm{total}}^{\mathrm{PF}}$ profile is smooth at time 0, even though some $V^{(i)}$ profiles discontinuously change at time 0 due to the adjacent-list update. Note that the $V_{\mathrm{total}}^{\mathrm{PF}}$ profile exhibits slight discontinuities at 50 fs intervals, which is due to the Gaussian hill deposition.

Concerning the bias forces as well, the $F^{(i)}$ profile of the migrating atom differs from those of the other atoms for the same reason. During the period before and after the list update, only the exponential weight, $w^{(i)}$, of the migrating atom becomes nearly equal to unity in contrast to the negligible weights of the other atoms. Consequently, the bias force acts only on the migrating atom, which is continuous at the list update because of the smooth $V^{(i)}$ profile of the migrating atom. See panels (c) and (d) in Fig. S1 showing the $F^{(i)}$ profile of each adjacent atom before and after the list update, in which no discontinuous force spikes act on any individual atom.

It is clear that the above discussion is held even when there are multiple vacancies or the presence of impurities. The CV reassignment occurs only when an adjacent atom in the transition state (migrating atom) crosses a Voronoi boundary. The total bias potential and bias forces are governed only by the migrating atom assigned to the same CV space. As a result, the continuities of both $V_{\mathrm{total}}^{\mathrm{PF}}$ and $F^{(i)}$ for any individual atom. We believe that our proposed approach is highly useful as a versatile tool for analyzing vacancy dynamics in various crystals.

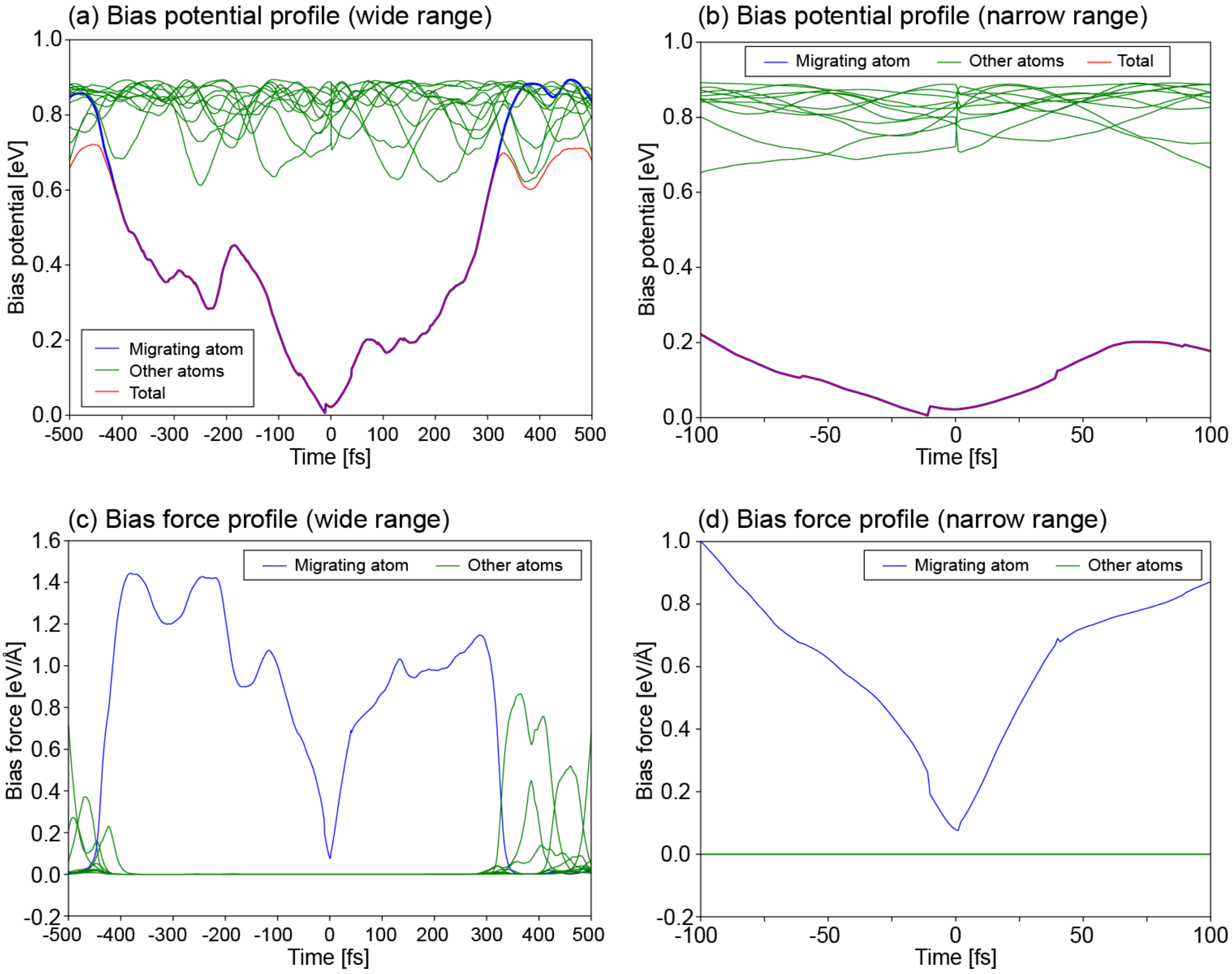

FIG. S1. (a)(b) The $V_{\text{total}}^{\text{PF}}$ profile (red line) and the individual $V^{(i)}$ profiles (Migrating atom: blue line, Other atoms: green lines) during 1000 fs and 200 fs before and after the adjacent-atom list update in a PB-MetaD simulation for the monovacancy diffusion in fcc-Cu at 500 K. The adjacent-atom list was updated at $t = 0$ fs. (c)(d) The $\|F^{(i)}\|$ profiles of the migrating atom (blue line) and other adjacent atoms (green lines) within the same time intervals in the same simulation.

Alt text: As an adjacent atom approaches the transition state near a Voronoi boundary, the total bias potential begins to coincide with the single $V^{(i)}$ profile of the migrating atom. The $V_{\text{total}}^{\text{PF}}$ profile is smooth at time 0, even though some $V^{(i)}$ profiles discontinuously change at time 0 due to the adjacent-list update. The bias force acts only on the migrating atom, which is also continuous at the list update because of the smooth $V^{(i)}$ profile of the migrating atom.

**S3. Convergence tests of MetaD simulations for setting parameters**

Since the convergence of MetaD simulations can depend on numerical parameters, the preliminary test calculations were performed by using the Cu monovacancy diffusion at 500 K as a model case. In the test calculations, the initial height, width, and deposition interval of the Gaussian hills were changed in the range of 0.02 – 0.08 eV, 0.2 – 0.8 Å, and 50 – 400 fs, respectively. Under every condition, eight independent simulations were performed to assess the statistical uncertainty.

Figure S2 shows the profiles of the free energy barrier for Cu monovacancy jumps in fcc Cu as a function of the hill deposition step, which correspond to the dependences on (a) the initial height, (b) width, and (c) deposition interval of Gaussian hills. The black line in each figure corresponds to the reference condition, under which the height, width, and deposition interval were set to 0.02 eV, 0.2 Å, and 50 fs, respectively. These profiles indicate that the MetaD convergence is considerably sensitive to the initial hill height. The free energy barrier tends to be overestimated in the early stage of the simulation, and this tendency is more pronounced as the initial hill height increases. By contrast, the convergence is relatively insensitive to the hill width, although the free energy barrier tends to be underestimated in the early stage of the simulation with the hill width excessively large (0.8 Å). Concerning the hill deposition interval, it does not significantly affect the MetaD convergence within the range of 50 – 400 fs.

Based on the test calculations, the smaller gaussian hill leads to the better convergence of MetaD simulations. Therefore, we employed smaller gaussian hill (initial height: 0.01 eV, width: 0.15 Å) with the short deposition interval (50 fs) in the accurate simulations to construct the vacancy FESs for the Cu monovacancy, divacancy, and In impurity diffusions in fcc Cu. For the O vacancy diffusion in rutile-$TiO_2$, we employed relatively large gaussian hill (initial height: 0.02 eV, width: 0.2 Å), because the potential energy barriers of O vacancy jumps are reported to be beyond 1 eV.

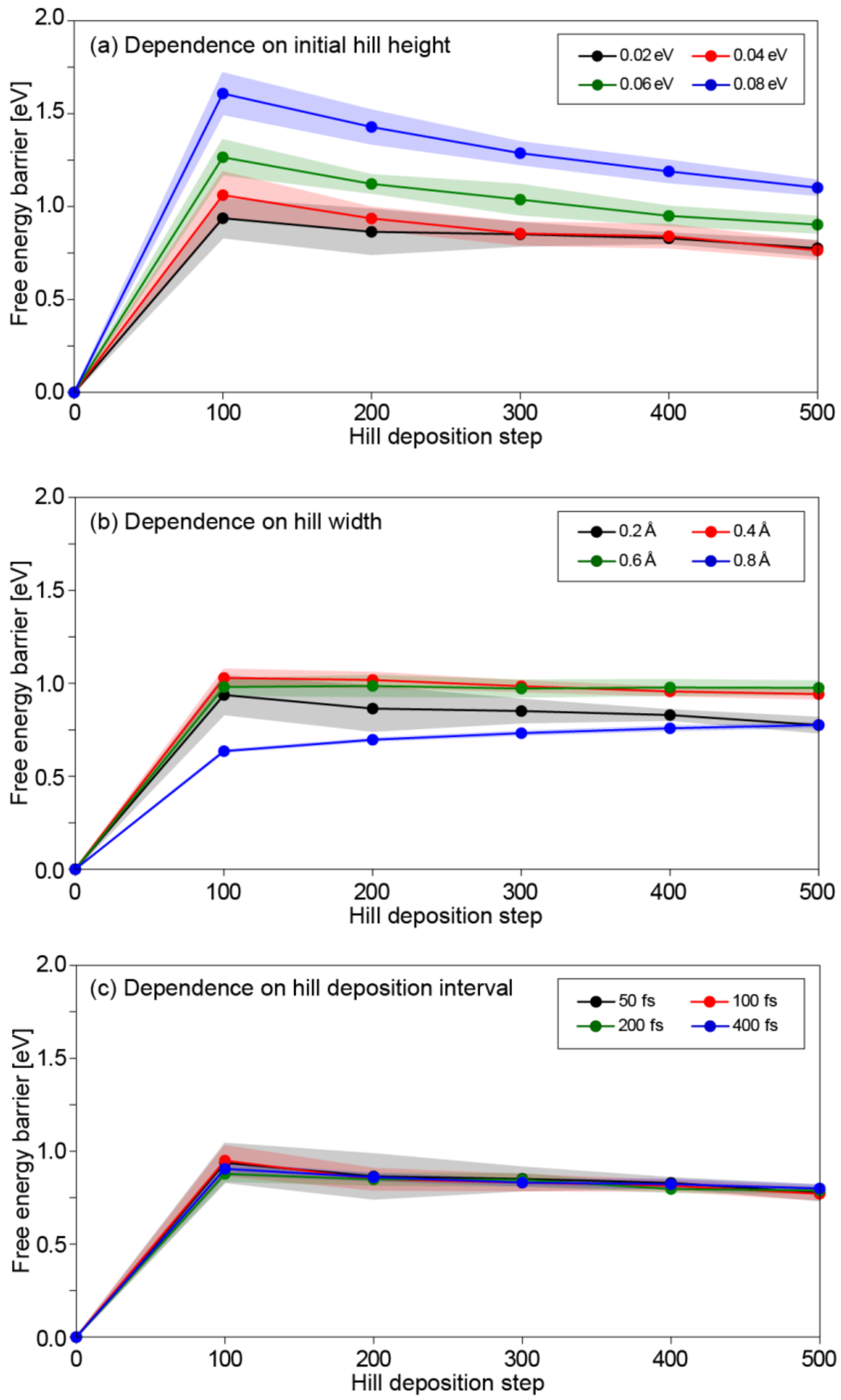

FIG. S2. The profiles of the free energy barrier for Cu monovacancy jumps in fcc Cu as a function of the hill deposition step, which correspond to the dependences on (a) the initial height, (b) width, and (c) deposition interval of Gaussian hills. The black line in each figure corresponds to the reference condition, under which the height, width, and deposition interval were set to 0.02 eV, 0.2 Å, and 50 fs, respectively. The solid lines and shaded regions denote the mean and $\pm\sigma$ over eight independent runs under each condition, respectively.

Alt text: (a) The free energy barrier tends to be overestimated in the early stage of the simulation, and this tendency is more pronounced as the initial hill height increases. (b) The free energy barrier profile is relatively insensitive to the hill width, although the free energy barrier tends to be underestimated in the early stage of the simulation with the hill width excessively large (0.8 Å). (c) Concerning the hill deposition interval, it does not significantly affect the free energy barrier profile within the range of 50 – 400 fs.

## S4. Accuracies of the constructed MLFFs

For the MD and MetaD simulations, the on-the-fly machine learning force fields (MLFFs) implemented in the VASP code were employed. The MLFFs were trained during the MD and preliminary PB-MetaDPF simulations using the on-the-fly learning scheme. The constructed MLFFs were subsequently employed for the accurate PB-MetaDPF simulations. Their accuracy was assessed by comparing total energies and atomic forces obtained from MLFFs and DFT calculations for several hundred structures sampled during the accurate PB-MetaDPF simulations (Figs. S3 and S4).

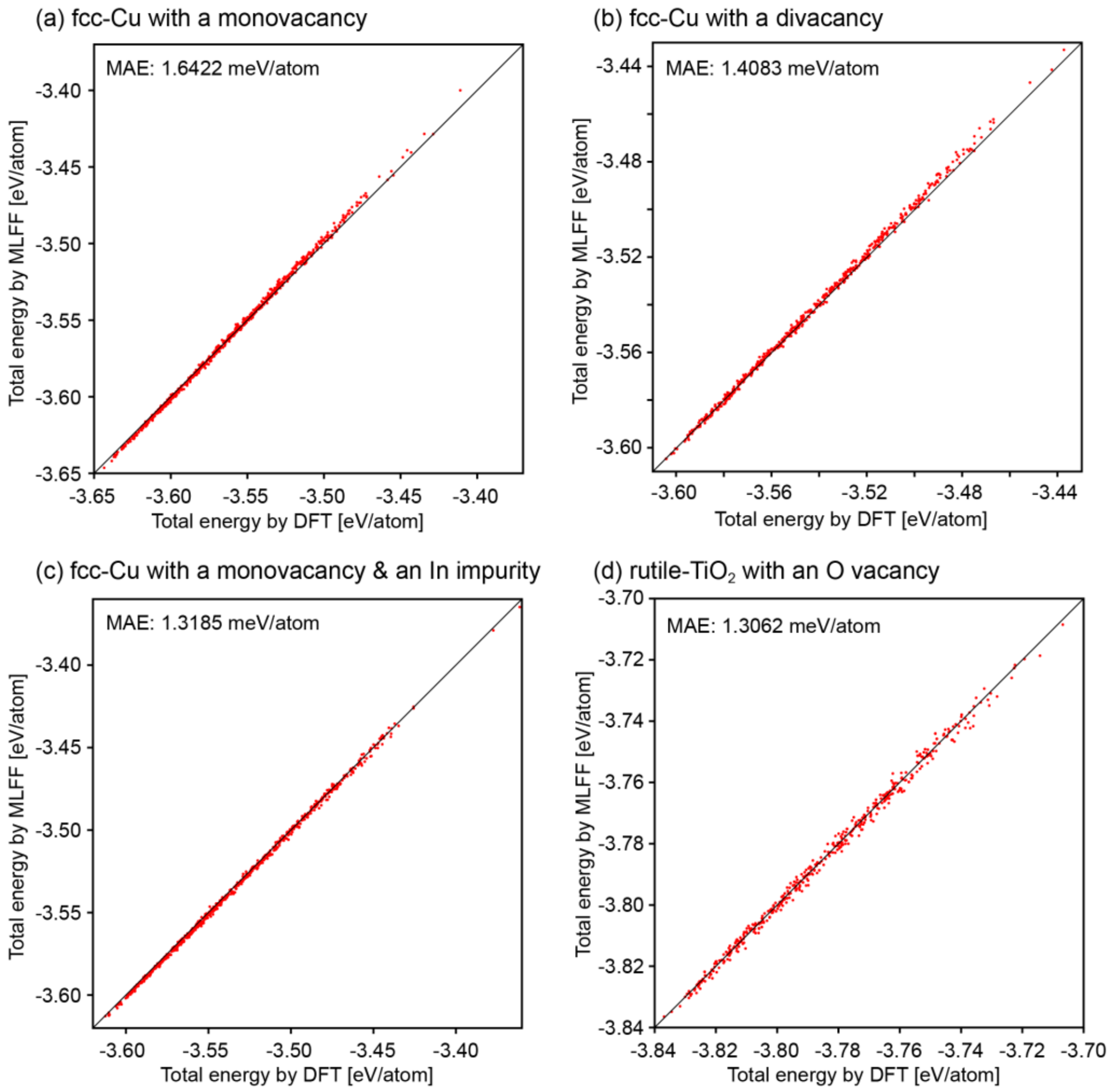

FIG. S3. Comparison of total energies obtained from the MLFFs and DFT calculations for several hundred structures sampled during the accurate PB-MetaDPF simulations: (a) fcc-Cu supercells with a monovacancy, (b) fcc-Cu supercells with a divacancy, (c) fcc-Cu supercells with a monovacancy and an In impurity, and (d) rutile-$TiO_2$ supercells with an O vacancy.

Alt text: In any system, the mean absolute error of the predicted total energy is less than 2 meV/atom, indicating that the constructed machine learning force fields have sufficient accuracies.

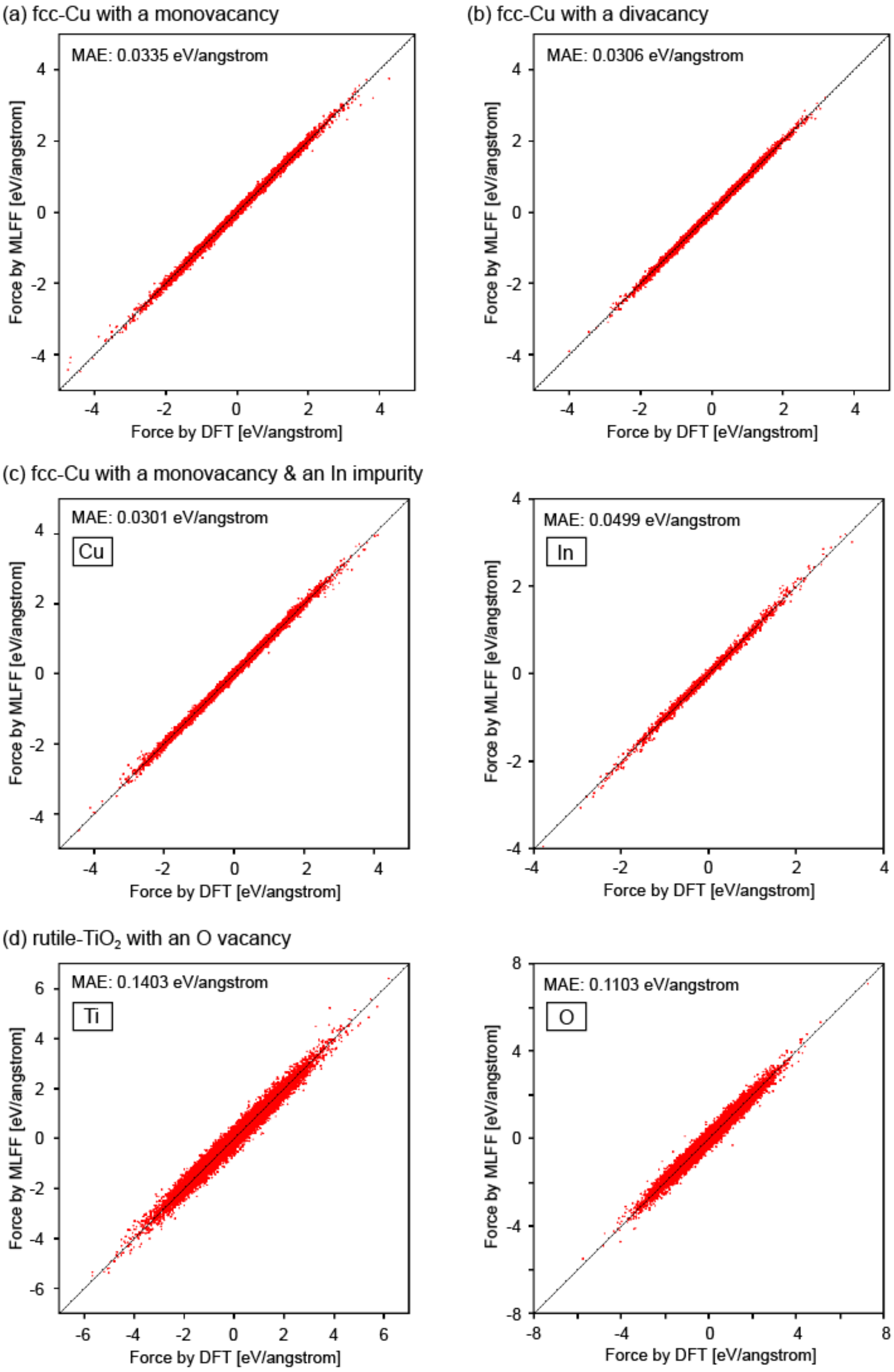

FIG. S4. Comparison of atomic forces obtained from the MLFFs and DFT calculations for several hundred structures sampled during the accurate PB-MetaDPF simulations: (a) fcc-Cu supercells with a monovacancy, (b) fcc-Cu supercells with a divacancy, (c) fcc-Cu supercells with a monovacancy and an In impurity, and (d) rutile-$TiO_2$ supercells with an O vacancy.

Alt text: For the fcc-Cu systems, the mean absolute errors of the predicted atomic forces are less than 0.05 eV/Å, while they are 0.14 eV/Å for Ti atoms and 0.11 eV/Å for O atoms in the rutile-$TiO_2$ systems.

**S5. Phonon calculations in fcc-Cu**

Phonon calculations were performed for fcc-Cu supercells containing a monovacancy to estimate vacancy jump frequencies under the harmonic approximation. The calculated vibrational eigenfrequencies at the initial and saddle-point states are shown in Fig. S5(a), where the region below 0 THz corresponds to imaginary frequencies. All vibrational modes exhibit real frequencies at the initial state, whereas one imaginary mode associated with the migration coordinate appears at the saddle-point state. The preexponential factor of the atomic jump frequency is the ratio of the products of vibrational eigenfrequencies at the initial and saddle-point states, excluding the imaginary mode at the saddle-point state (see Eq. (2) in the main text).

Additional phonon calculations were performed for a perfect supercell and a defective supercell containing a divacancy to estimate the site fractions $x$ of monovacancies and divacancies in fcc-Cu under the harmonic approximation. The defect formation free energies of a monovacancy and a divacancy, $\Delta F_{\alpha}^{\mathrm{form}}(T)$ ($\alpha = \mathrm{v}, 2\mathrm{v}$), respectively, are expressed as

$$\Delta F_{\alpha}^{\mathrm{form}}(T) = (E_{\alpha}^{\mathrm{total}} + F_{\alpha}^{\mathrm{vib}}(T)) - (E_{\mathrm{perfect}}^{\mathrm{total}} + F_{\mathrm{perfect}}^{\mathrm{vib}}(T)) \times \frac{n_{\alpha}}{n_{\mathrm{perfect}}},$$

where $T$ is the temperature, and $E_{\alpha}^{\mathrm{total}}$ and $E_{\mathrm{perfect}}^{\mathrm{total}}$ are the total energies of the defective and perfect supercells, respectively. $F_{\alpha}^{\mathrm{vib}}(T)$ and $F_{\mathrm{perfect}}^{\mathrm{vib}}(T)$ are the vibrational free energies, and $n_{\alpha}$ and $n_{\mathrm{perfect}}$ are the number of atoms in the defective and perfect supercells, respectively. The vibrational free energies, $F^{\mathrm{vib}}(T)$, were estimated under the harmonic approximation as

$$F^{\mathrm{vib}}(T) = \sum_i \frac{h\nu_i}{2} + k_{\mathrm{B}}T \sum_i \ln[1 - \exp(-h\nu_i/k_{\mathrm{B}}T)],$$

where $h$ and $k_{\mathrm{B}}$ are the Planck and the Boltzmann constants, respectively, and $\nu_i$ is the $i$th vibrational eigenfrequency. The $q$-point mesh in the Brillouin zone was set to 15×15×15 for the supercells. The site fraction, $x_{\alpha}$, is given by

$$x_{\alpha}(T) = n_{\alpha}^{\mathrm{config}} \exp\left(-\Delta F_{\alpha}^{\mathrm{form}}(T)/k_{\mathrm{B}}T\right),$$

where $n_{\alpha}^{\mathrm{config}}$ is the number of defect configurations per lattice site, i.e., $n_{\mathrm{v}}^{\mathrm{config}} = 1$ and $n_{2\mathrm{v}}^{\mathrm{config}} = 6$ in fcc-Cu. Figure S5(b) shows the estimated site fractions of monovacancies and divacancies in fcc-Cu as functions of inverse temperature. The apparent activation energies of site fractions are 1.11 eV for monovacancy and 2.16 eV for divacancy, respectively. The slope in the Arrhenius plots of Cu tracer diffusion coefficients via monovacancy and divacancy mechanisms are approximately equal to the sum of slopes for the vacancy jump frequency and site fraction.

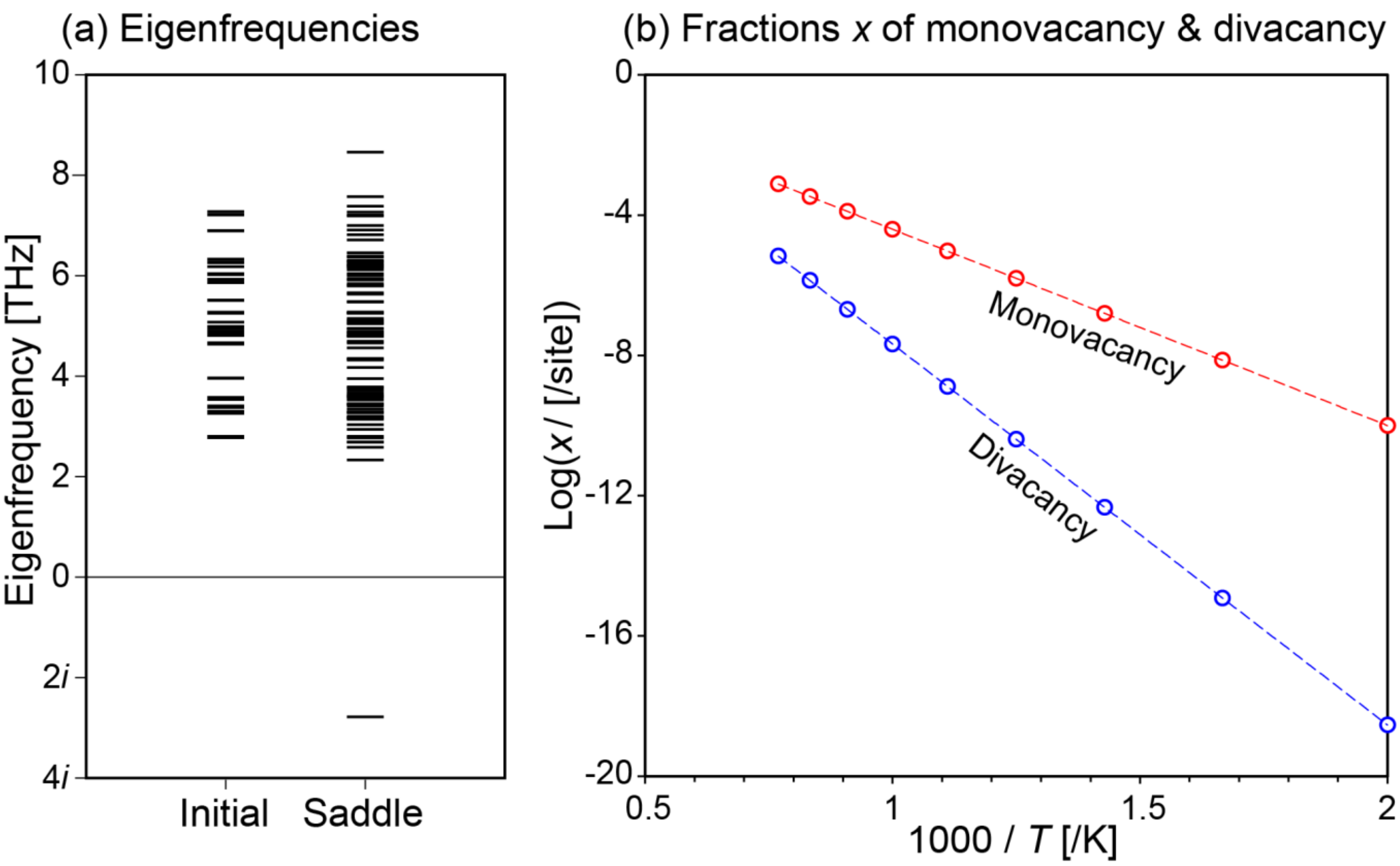

FIG. S5. (a) Calculated vibrational eigenfrequencies at the initial and saddle-point states for a monovacancy jump in fcc-Cu. The region below 0 THz corresponds to imaginary frequencies. (b) Estimated site fractions of monovacancies and divacancies in fcc-Cu, shown by red and blue symbols, respectively.
Alt text: (a) All vibrational modes exhibit real frequencies at the initial state, whereas one imaginary mode associated with the migration coordinate appears at the saddle-point state. (b) The apparent activation energies of site fractions are 1.11 eV for monovacancy and 2.16 eV for divacancy, respectively.

## S6. One-dimensional FESs

Figure S6 shows the one-dimensional FESs along the migration coordinate obtained by the PB-MetaDPF simulations at 500, 700, 900, 1100, and 1300 K. For comparison, the black broken lines are the free energy curves estimated under the harmonic approximation. The one-dimensional FESs indicate that the curvatures around the free energy minima are smaller than those under the harmonic approximation and does not change significantly with temperature. Concerning the curvature around the free energy maxima, it coincides with that under the harmonic approximation at 500 K, but gradually decreases and deviates from the harmonic approximation curve with temperature. Thus, even when focusing only on the direction along the migration coordinate, the anharmonicity is not neglected, which leads to the significant deviation, particularly around the saddle point at high temperatures. This result suggests that it is necessary to take anharmonicity into account for rigorously estimating atomic jump frequencies based on the transition state theory.

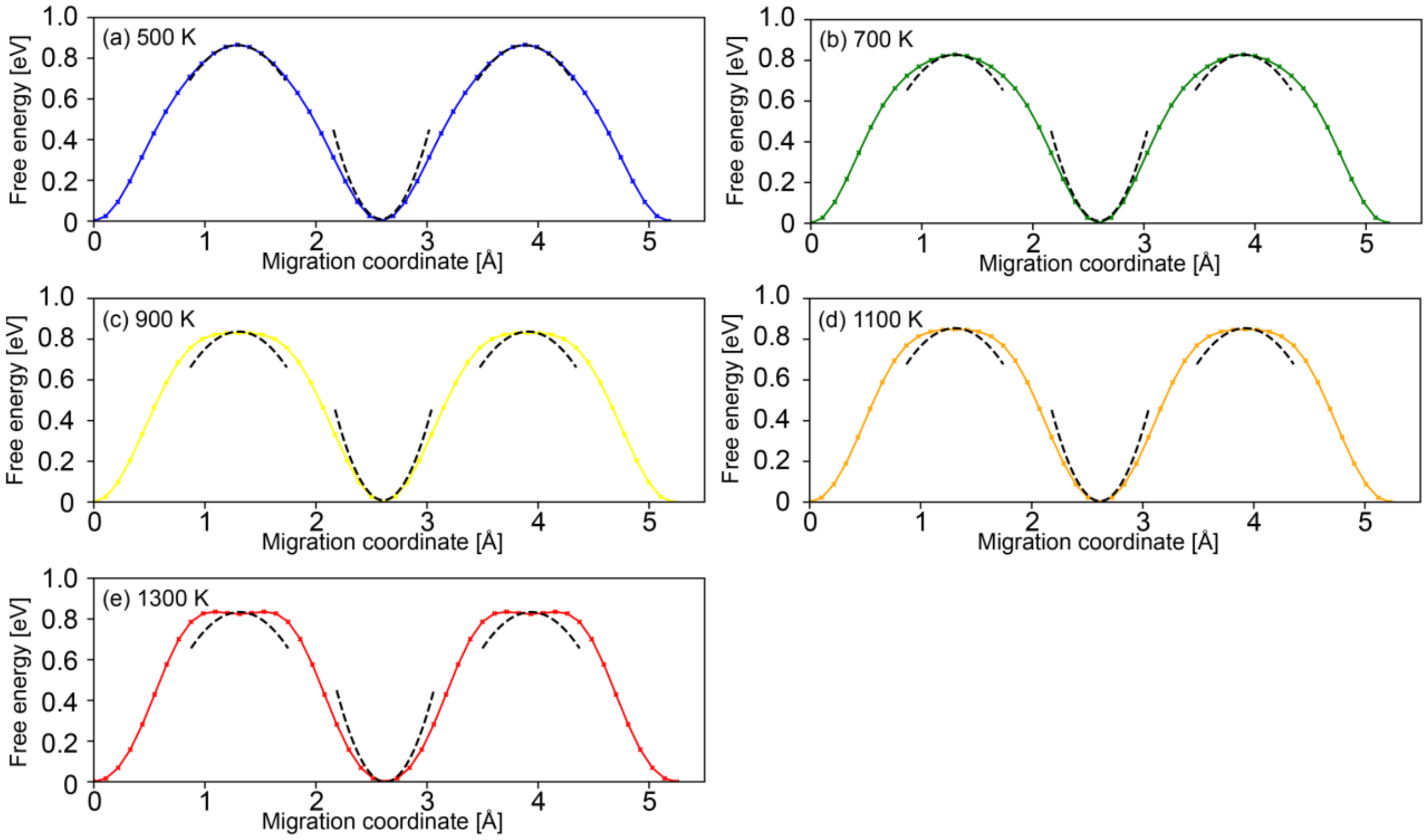

FIG. S6. One-dimensional FESs along the migration coordinate obtained by the PB-MetaDPF simulations at (a) 500 K, (b) 700 K, (c) 900 K, (d) 1100 K, and (e) 1300 K. The black broken lines are the free energy curves estimated under the harmonic approximation.
Alt text: The curvature around the free energy minimum is smaller than that under the harmonic approximation, and does not change significantly with temperature. Concerning the curvature around the free energy maximum, it coincides with that under the harmonic approximation at 500 K, but gradually decreases and deviates from the harmonic approximation curve with temperature.